%% file: Main.tex
\documentclass[a4paper,twoside]{article}

\usepackage{epsfig}
\usepackage{subcaption}
\usepackage{calc}
\usepackage{amssymb}
\usepackage{amstext}
\usepackage{amsmath}
\usepackage{amsthm}
\usepackage{multicol}
\usepackage{pslatex}
\usepackage{apalike}
\usepackage{algorithm2e}
\usepackage[most]{tcolorbox}
\usepackage[bottom]{footmisc}
\usepackage{graphicx}
\usepackage{url}
\usepackage{other/SCITEPRESS}     

\newcommand\blfootnote[1]{%
  \begingroup
  \renewcommand\thefootnote{}\footnote{#1}%
  \addtocounter{footnote}{-1}%
  \endgroup
}
\begin{document}

\graphicspath{ {./images/} }

\title{Applying Informer for Option Pricing: A Transformer-Based Approach}

\author{\authorname{}
\affiliation{}
}
\author{\authorname{Feliks Bańka\sup{1}\orcidAuthor{0009-0005-1973-5861}, Jarosław A. Chudziak\sup{1}\orcidAuthor{0000-0003-4534-8652}}
\affiliation{\sup{1}The Faculty of Electronics and Information Technology, Warsaw University of Technology, Poland}
}


\abstract{
Accurate option pricing is essential for effective trading and risk management in financial markets, yet it remains challenging due to market volatility and the limitations of traditional models like Black-Scholes. In this paper, we investigate the application of the Informer neural network for option pricing, leveraging its ability to capture long-term dependencies and dynamically adjust to market fluctuations. This research contributes to the field of financial forecasting by introducing Informer's efficient architecture to enhance prediction accuracy and provide a more adaptable and resilient framework compared to existing methods. Our results demonstrate that Informer outperforms traditional approaches in option pricing, advancing the capabilities of data-driven financial forecasting in this domain.
}

\onecolumn
\maketitle
\normalsize
\setcounter{footnote}{0}

\input{introduction}
\input{literature}
\input{architecture}

\input{experiments}
\input{Conclusion}

\bibliographystyle{apalike}
{\small
\bibliography{Main}}

\end{document}

%% file: introduction.tex
\section{Introduction}
Option pricing is a cornerstone of modern finance, essential for developing trading strategies and managing risk. Options enable traders and investors to hedge against potential losses or speculate on price movements. A call (put) option grants the holder the right, but not the obligation, to buy (sell) an asset at a specified price before the contract expires. Accurate option pricing models shape critical decisions in hedging and risk management, directly affecting trading portfolio profitability and stability.
\blfootnote{\noindent \hspace{-0.65cm}\hrulefill\\ This is the accepted version of the paper presented at the \textbf{17th International Conference on Agents and Artificial Intelligence} (ICAART 2025), Porto, Portugal. Available at:\\\url{https://doi.org/10.5220/0013320900003890}}

Early theoretical frameworks, such as the Black–Scholes~\cite{BlackMyron,Merton1973} and the Heston~\cite{Heston1993} models, offered valuable mathematical foundations but often rely on simplifying assumptions (e.g., constant volatility). These assumptions do not always hold in real-world markets, where sudden shifts in macroeconomic conditions or sentiment can lead to rapid changes in asset prices~\cite{Bollerslev1986}. Over the past few decades, machine learning techniques—such as LSTM-based neural networks~\cite{Hochreiter1997,Liu2023,Bao2017}—have demonstrated improved adaptability by capturing non-linearities and sequential dependencies. Yet, their effectiveness can be limited when handling very long time sequences, which demand more efficient and robust architectures.

Transformer-based models, originally devised for natural language processing~\cite{Vaswani2017}, have shown promise in overcoming these challenges by leveraging self-attention mechanisms that allow for parallelized long-sequence processing. Recent advances, such as the Informer model~\cite{Zhou2021}, have introduced more efficient attention mechanisms geared toward time-series data. However, their application within option pricing remains underexplored, motivating the present study to investigate whether Informer’s long-horizon capability and computational efficiency can produce more accurate predictions in option pricing tasks.

This paper contributes to the field of financial modeling by evaluating the application of the Informer architecture for predicting option prices, leveraging its efficient attention mechanism and long-sequence modeling capabilities to enhance prediction accuracy and adaptability to market fluctuations. Informer's ability to handle long-term dependencies makes it an ideal candidate for modeling complex financial data, offering a more advanced approach compared to traditional models like Black-Scholes \cite{BlackMyron,Merton1973} and Heston \cite{Heston1993}, as well as existing machine learning models such as LSTM \cite{Hochreiter1997,Liu2023}.
The contributions of this paper are as follows:
\begin{itemize}
    \item We apply the Informer architecture to option pricing, leveraging its long-sequence modeling capabilities and self-attention mechanisms to enhance prediction accuracy.
    \item We benchmark the model against traditional and machine learning-based approaches, evaluating its performance in high-volatility scenarios.
    \item We present an analysis of Informer's predictive accuracy and trading profitability on historical data.
\end{itemize}

The remainder of this paper is organized as follows: Section 2 discusses related work, focusing on traditional and machine learning approaches to option pricing and the emerging role of Transformers in finance. Section 3 outlines the Informer-based methodology applied to option pricing. Section 4 presents the experimental setup and results, and Section 5 concludes with a summary and potential directions for future research.

%% file: literature.tex
\section{Related Work}
The foundational models for option pricing, such as the Black-Scholes model~\cite{BlackMyron,Merton1973} and the binomial model~\cite{Cox1979}, have been pivotal in shaping early financial derivatives pricing. These models introduced critical concepts such as risk-neutral valuation but often rest on simplifying assumptions, such as constant volatility, which do not align with real-world market conditions. The introduction of stochastic volatility models, such as the Heston model~\cite{Heston1993}, offered more flexibility by allowing volatility to vary as a stochastic process.

Despite improvements like stochastic volatility in the Heston model~\cite{Heston1993}, traditional models remain limited in capturing the rapid shifts and complex dependencies of modern financial markets~\cite{Jones2019,Eisdorfer2022}. This has motivated the exploration of adaptive machine-learning approaches capable of modeling intricate relationships and dynamic patterns in financial data~\cite{Gatheral2006,Christoffersen2009}.

Recurrent architectures, such as Long Short-Term Memory (LSTM) networks and Gated Recurrent Units (GRU), became popular due to their ability to capture temporal dependencies in sequential data~\cite{MINTARYA202396,Hochreiter1997,Liu2023}. However, these models encounter scalability challenges when dealing with long-term dependencies or high-frequency data, often leading to computational inefficiencies~\cite{Binkowski2018,Lim2019}. While modular and hybrid neural networks have been employed to integrate financial indicators and better capture non-linearities, issues of scalability and interpretability persist~\cite{Amilon2003,Gradojevic2009}.

\begin{figure}[h!]
\centering
\includegraphics[width=0.41\textwidth]{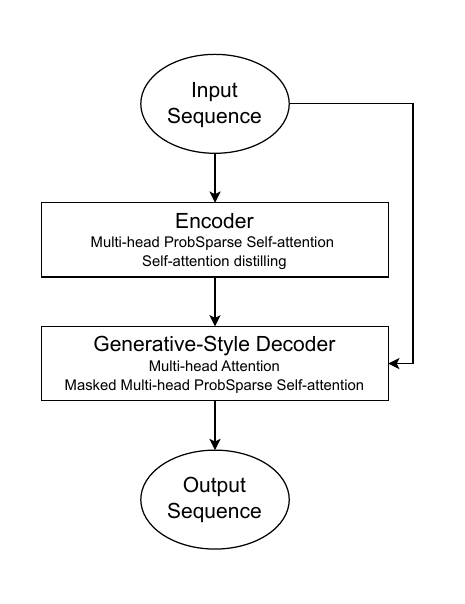}
\caption{Informer model - concepcual. Based on~\cite{Szydlowski2024}.}
\label{fig:informer_overview}
\end{figure}

Transformers, initially developed for natural language processing~\cite{Vaswani2017}, introduced self-attention mechanisms that bypass the limitations of recurrent models, allowing for the capture of long-term dependencies without the vanishing gradient problem. Szydlowski~\cite{Szydlowski2024b,Wawer2024} applied the Hidformer model to stock market prediction, demonstrating its effectiveness in handling long sequences and capturing complex market patterns. Informer, introduced by Zhou et al.~\cite{Zhou2021} and illustrated in Figure~\ref{fig:informer_overview}, marked a significant advancement for time-series analysis with its ProbSparse self-attention mechanism, reducing the time and memory complexity of processing long sequences to $O(L~\log L)$ for input length $L$.
Wang et al.~\cite{Wang2022} demonstrated Informer's application in predicting stock market indices, showcasing its ability to outperform traditional deep learning models (e.g., CNN, RNN, LSTM) by effectively capturing relevant information while filtering out noise—a common challenge in financial time series. Informer's robust multi-head attention mechanism allowed for the extraction of key features, leading to significantly higher prediction accuracy, particularly in short-term forecasting.

While studies have applied Transformer-based architectures to option pricing, including the generic Transformer model used by Guo and Tian~\cite{Guo2022} and Sagen's investigation of the Temporal Fusion Transformer (TFT)~\cite{Sagen2024}, the application of Informer has not been explored in this domain. Given Informer's strengths in long-sequence modeling and handling high-dimensional data efficiently, this paper seeks to evaluate its potential for enhancing predictive accuracy and computational efficiency in the complex landscape of option pricing.

%% file: architecture.tex
\section{Model Architecture}

In this section, we outline the architecture of the Informer-based model employed for option pricing. The Informer model is chosen for its ability to handle long sequences efficiently and capture dependencies over varying time scales through its unique attention mechanisms and architectural optimizations~\cite{Zhou2021,Wang2022}. This is essential in financial applications where complex temporal relationships can influence outcomes significantly.

\begin{figure}[h!]
\centering
\includegraphics[width=0.5\textwidth]{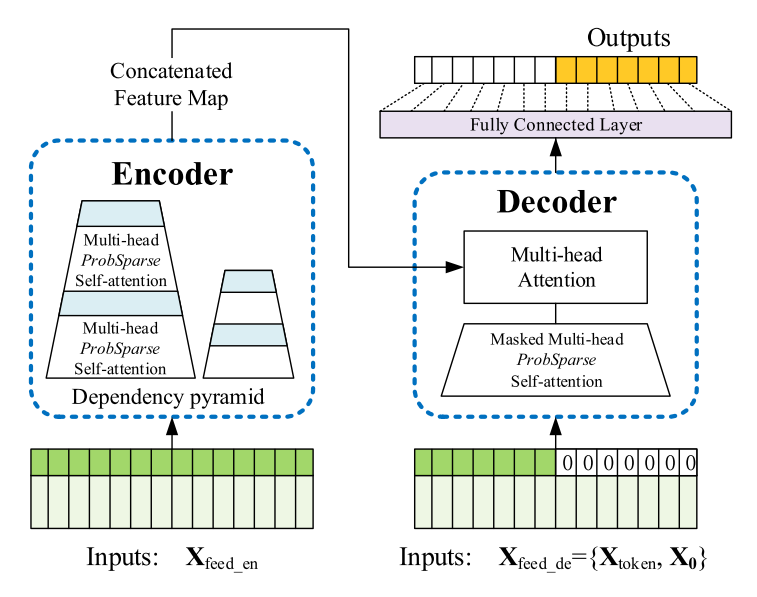}
\caption{Informer model overview. Copied from~\cite{Zhou2021}.}
\label{fig:informer_overview}
\end{figure}

\subsection{Data and Feature Engineering}
For effective model training, we select input features known to be crucial for option pricing. These features include the underlying asset price, implied volatility, time to maturity, strike price, and an indicator for the type of option (call or put). Each of these factors provides valuable insights into how option prices respond to market conditions. Volatility measures indicate market uncertainty~\cite{Hull06,Kolm2019}, while time to maturity and strike price are fundamental in assessing the intrinsic and extrinsic value of the option~\cite{Jones2019,BlackMyron,Merton1973}.
Normalization is applied to standardize the data, ensuring all features are on a comparable scale:
\begin{equation}
    x_t^{\text{norm}} = \frac{x_t - x_{\text{min}}}{x_{\text{max}} - x_{\text{min}}}
\end{equation}
where $x_t^{\text{norm}}$ represents the normalized feature value at time $t$, and $x_{\text{max}}$ and $x_{\text{min}}$ denote the maximum and minimum feature values, respectively. This approach keeps all features within the range $[0, 1]$, aiding in model stability and faster convergence during training.

The input data is structured as a time-series sequence with a moving window approach, where $T_x$ past data points are used to predict $T_y$ future option prices or metrics. This sequential setup helps capture dependencies over different time horizons and enables the model to account for short-term fluctuations as well as long-term trends.

\subsection{Proposed Model Architecture}

The Informer-based model extends the standard Transformer architecture by incorporating enhancements tailored to the challenges of time-series forecasting in financial applications. It consists of two main components - the encoder and the decoder, which exchange information through self-attention mechanisms and encoder-decoder attention modules, as we can see in Figure~\ref{fig:informer_overview}. This section details each of these components and the overall data flow and token construction procedure.

\subsubsection{Encoder}

The encoder is responsible for extracting meaningful temporal dependencies from the input sequence. It includes an embedding layer, a ProbSparse self-attention mechanism, a feedforward sub-layer, and a self-attention distilling step to reduce computational overhead.

\textbf{Embedding Layer.}
Each time step in the raw data is represented as a \emph{token}, which is a set of features (e.g., strike price, time to maturity). The embedding layer projects these tokens into a dense vector space of fixed dimension, enabling the network to learn hidden interactions across features.

\begin{figure}[h!]
\centering
\includegraphics[width=0.5\textwidth]{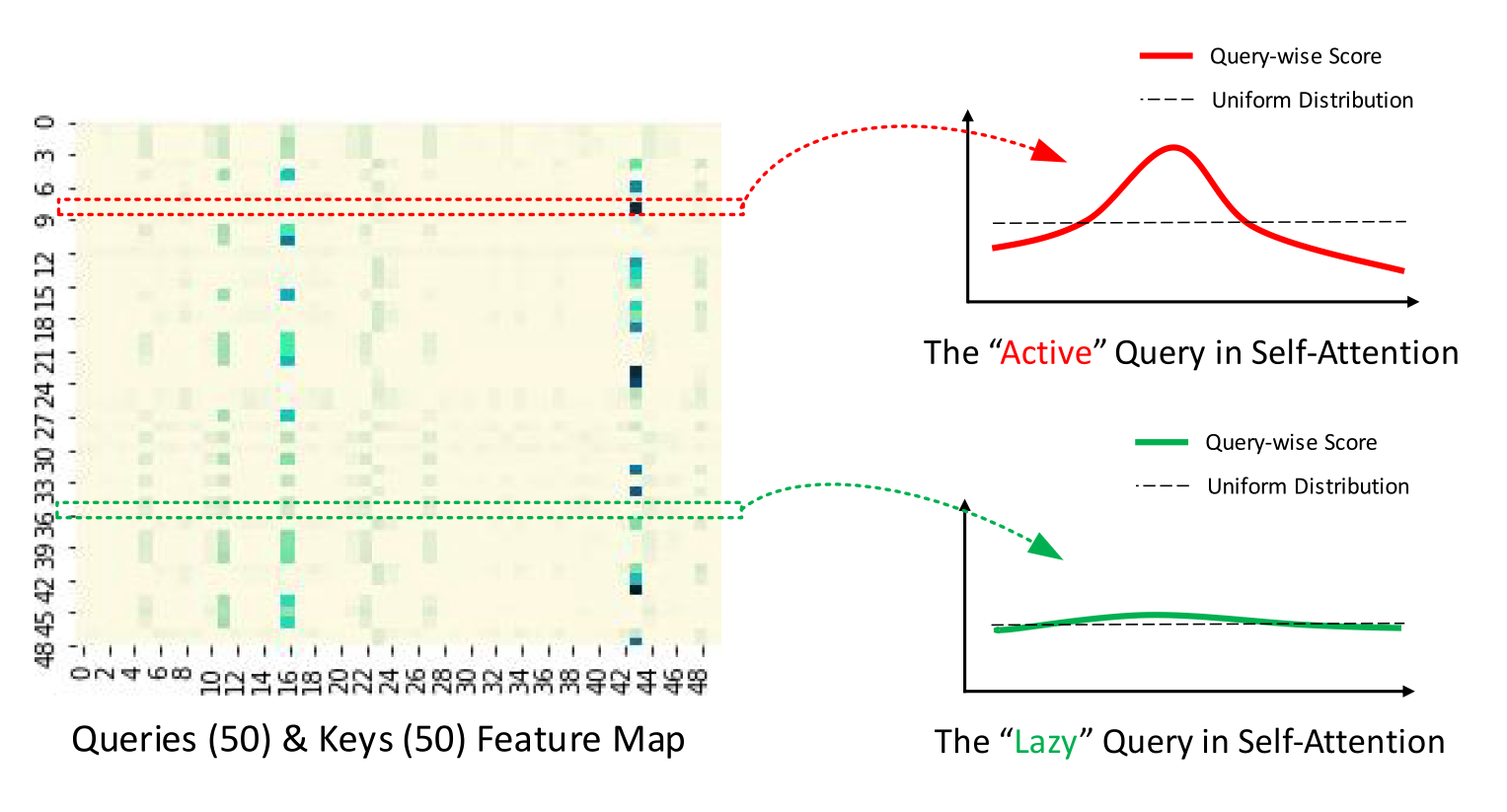}
\caption{Illustration of the ProbSparse Attention mechanism. Adapted from~\cite{Zhou2021}.}
\label{fig:probsparse_attention}
\end{figure}

\textbf{ProbSparse Self-Attention Mechanism.}
This attention mechanism aims to identify and focus on the most informative queries in the attention calculation, as illustrated in Figure~\ref{fig:probsparse_attention}. Instead of computing attention scores for all $L$ queries and keys, it selects a subset of queries based on the \emph{Kullback-Leibler divergence} (KLD) between the query distribution and a predefined sparse distribution. Formally:

\begin{equation}
    \text{Attention}(Q, K, V) = \text{Softmax}\left( \frac{Q^\top K}{\sqrt{d_k}} \right) V
\end{equation}
where $Q, K, V$ are the query, key, and value matrices, and $d_k$ is the dimension of the keys. By selecting only the top-$U$ queries (with $U \ll L$), complexity is reduced from $\mathcal{O}(L^2)$ to approximately $\mathcal{O}(L \log L)$, making the model scalable for long sequences.

\textbf{Feedforward Sub-Layer.}
The output of the attention sub-layer is passed through a fully connected feedforward network with a hidden dimensionality \(D_{\text{FF}}\):
\begin{equation}
    \text{FeedForward}(x) = \text{ReLU}(W_1 x + b_1)W_2 + b_2,
\end{equation}
where \(W_1, W_2\) are weight matrices, and \(b_1, b_2\) are biases. A larger dimension \(D_{\text{FF}}\) allows the model to capture intricate patterns.

\textbf{Self-Attention Distilling.}
To improve efficiency, the Informer applies a distilling mechanism at the end of each layer, pooling or downsampling the sequence to retain only the most critical tokens. Mathematically,
\[
    \mathbf{Z}^l \;=\; \mathrm{Pooling}(\mathbf{X}^l), 
    \quad
    \mathbf{X}^{l+1} \;=\; \mathrm{SelfAttention}(\mathbf{Z}^l),
\]
where $\mathbf{X}^l$ is the layer-$l$ input. This process concentrates the model’s capacity on dominant features, enhancing performance while mitigating overfitting.

\textbf{Encoder Output.}
The final encoder output, denoted by $\mathbf{E_t}$, is a contextually enriched representation of the input tokens and will be passed to the decoder for generating forecasts.

\subsubsection{Decoder}

The decoder produces the target sequence by leveraging both the encoder output and partially known future labels. It features a self-attention sub-layer, encoder-decoder attention, and a feedforward network. Unlike the traditional approach, which decodes one step at a time, the Informer employs a generative-style approach to predict all future steps simultaneously.

\textbf{Decoder Input Preparation.}
To provide the decoder with partial knowledge of the future horizon, the model concatenates the most recent $T_{\mathrm{label}}$ observed values with placeholder zeros for the $T_y$ unknown time steps. This can be expressed as:
\[
    \mathbf{D}_t 
    \;=\;
    [\,y_{t - T_{\mathrm{label}} + 1},\,\ldots,\,y_t,\,0,\,\ldots,\,0\,].
\]
During training, the first portion corresponds to known labels or ground truth values, while zeros mark positions to be predicted.

\textbf{Attention Modules and Feedforward Sub-Layer.}
In the decoder, self-attention accounts for dependencies among known and future positions in $\mathbf{D}_t$, while encoder-decoder attention utilizes $\mathbf{E_t}$ (the encoder output) as keys and values to incorporate previously extracted temporal structure. It also applies a feedforward sub-layer similar to that in the encoder.

\textbf{Generative-Style Decoding.}
Finally, the decoder produces the entire predicted sequence in one forward pass:
\[
    \hat{\mathbf{Y}}_t \;=\; \mathrm{Decoder}(\mathbf{E}_t,\,\mathbf{D}_t).
\]
This approach improves inference speed compared to autoregressive decoding, which is advantageous for time-sensitive financial applications.

\textbf{Decoder Output.}
The vector $\hat{\mathbf{Y}}_t$ constitutes the model’s forecast for the $T_y$ future time steps.

\subsection{Model Workflow}

The overall workflow begins by converting each time step into a token that bundles relevant features. These tokens are then passed to the embedding layer, which maps them into a continuous space of dimension $d_{\mathrm{model}}$. The encoder applies ProbSparse self-attention, feedforward transformations, and self-attention distilling to capture critical dependencies with reduced computational overhead. Its final output $\mathbf{E_t}$, enriched with temporal context, is transferred to the decoder.

In parallel, the decoder constructs its input $\mathbf{D}_t$ by combining partially known labels from the prediction window with placeholder zeros. Self-attention in the decoder identifies dependencies among these elements, while encoder-decoder attention integrates signals from $\mathbf{E_t}$. The generative-style decoding step then yields a full multi-step forecast in a single pass, producing $\hat{\mathbf{Y}}_t$. This hierarchical design is especially suited to financial time-series forecasting, where long-range dependencies and efficient computation are both critical.

%% file: experiments.tex
\section{Experiments}

The experiments conducted aim to evaluate the robustness and predictive power of the proposed Informer-based model in the context of option pricing. A thorough comparison is established using baseline models that encompass traditional and machine learning-based methods.

\subsection{Dataset and Data Preparation}
We use a dataset comprising eight years of historical option contracts for Apple Inc. (AAPL), sourced from publicly available financial databases, covering the period from January 4, 2016, to March 31, 2023. The dataset includes both call and put options with varying strike prices, expiration dates, and moneyness levels, providing a diverse and comprehensive foundation for analysis.

To improve data quality and ensure relevance, the preprocessing stage included the application of strict selection criteria. Options with a time-to-maturity (TTM) below 30 days were excluded, as such short-term contracts are typically highly volatile and speculative~\cite{Heston1993}. Furthermore, only options with a moneyness ratio (the ratio of the underlying asset's price to the strike price) between 0.6 and 1.3 were included, as near-the-money options are more liquid and exhibit more reliable pricing~\cite{Bakshi2000}. Contracts with insufficient data points or low trading volume were also removed to maintain robustness and integrity.
The dataset is split into training, validation, and test sets, with 70\% of the data allocated for training, 15\% for validation, and the remaining 15\% for testing~\cite{Matsunaga2019}. This split ensures that the model is evaluated on unseen data, simulating real-world conditions where future predictions depend on past training.

\subsection{Model Configuration and Training Strategy}
The Informer model is configured to handle complex time-series data with the following parameters: the input sequence length is set to 30 days ($T_x = 30$), and predictions are made over a 30-day horizon ($T_y = 30$). The architecture includes one encoder layer and two decoder layers with  a label length of 5 days, each featuring three attention heads. The embedding dimension ($D_{\text{MODEL}}$) is set to 32, balancing computational efficiency and model expressiveness.
The feedforward network dimension is set to 8, with a dropout rate of 0.06 to prevent overfitting. The model employs full attention with a factor of 3, suitable for capturing temporal patterns effectively in financial time-series data.
The training process employs a batch size of 64 and utilizes the Adam optimizer~\cite{Kingma2014} with an initial learning rate of 0.0001. Training proceeds over 300 epochs, with early stopping applied based on validation loss, using a patience of 30 epochs. A weighted mean squared error (MSE) loss function is used, prioritizing accuracy across the entire 30-day prediction horizon.
Hyperparameters, including the number of layers, attention heads, embedding dimension, learning rate, and dropout rate, were fine-tuned via random search.


\subsection{Evaluation Metrics}
The performance of the Informer model is evaluated using a comprehensive set of metrics to ensure a robust evaluation~\cite{ruf2020neuralnetworksoptionpricing}:

\textbf{Prediction Accuracy:} The model's outputs are compared with the ground truth on the validation set to evaluate the prediction accuracy. Two commonly used indicators are employed: Mean Absolute Error (MAE), which measures the average magnitude of prediction errors, and Root Mean Squared Error (RMSE), which emphasizes larger errors to capture prediction variance. Lower values of both metrics indicate better model performance.

\textbf{Final-Day Evaluation:} We focus on final-day evaluation because it highlights the model's ability to make accurate long-term predictions, which is crucial for strategic financial decision-making~\cite{Kristoufek2012}. To measure this, we use Direction Accuracy (DA), which measures the percentage of sequences where the predicted and actual price changes have the same direction, and Final-Day MAE, which calculates the MAE between predicted and actual prices specifically on the last day.

\textbf{Return Calculation:} The trading effectiveness of the model is evaluated using a simple strategy based on the predicted price at the end of each sequence. For a given sequence, if the predicted price ($\hat{y}_{t+30}$) is higher than the starting price ($y_t$), a long position is taken; otherwise, a short position is assumed. The return for the sequence is calculated as: 
\begin{equation} R = \ln \frac{y_{t+30}}{y_t} \times \text{sign}(\hat{y}_{t+30} - y_t) \end{equation} 
where $y_{t+30}$ is the true price at the prediction horizon, $y_t$ is the starting price, and $\hat{y}_{t+30}$ is the predicted price.

The cumulative net value (NV) aggregates returns across all sequences in the dataset, starting from an initial value of 1: 
\begin{equation} NV = 1 + \sum_{i=1}^{N} R_i \end{equation} 
where $N$ is the total number of sequences.

By combining predictive accuracy metrics (MAE and RMSE) with trading performance (NV), this evaluation framework captures both the statistical precision and the practical utility of the model in financial applications.

To benchmark the performance of the Informer-based model, we compare it against several established baseline models, including the Black-Scholes model, the Heston model, and the simple LSTM-based model. These models, ranging from traditional finance to advanced machine learning, help evaluate how the Informer performs in option pricing, highlighting its strengths and areas for improvement.

\subsection{Results and Analysis}

The results of the experiments demonstrate that the Informer model consistently outperforms all other models, both in terms of prediction accuracy and final-day evaluation metrics. 

\begin{table}[h!]
\centering
\caption{Overall prediction metrics for all models.}
\begin{tabular}{|c|c|c|}
\hline
\textbf{Model} & \textbf{MAE} & \textbf{RMSE} \\
\hline
Informer & 2.7145 & 3.6766 \\
LSTM & 3.9343 & 5.0373 \\
Black-Scholes & 4.1765 & 5.3840 \\
Heston & 4.1282 & 5.3565 \\
\hline
\end{tabular}
\label{tab:overall_metrics}
\end{table}

\begin{figure}[h!]
\centering
\includegraphics[width=0.48\textwidth]{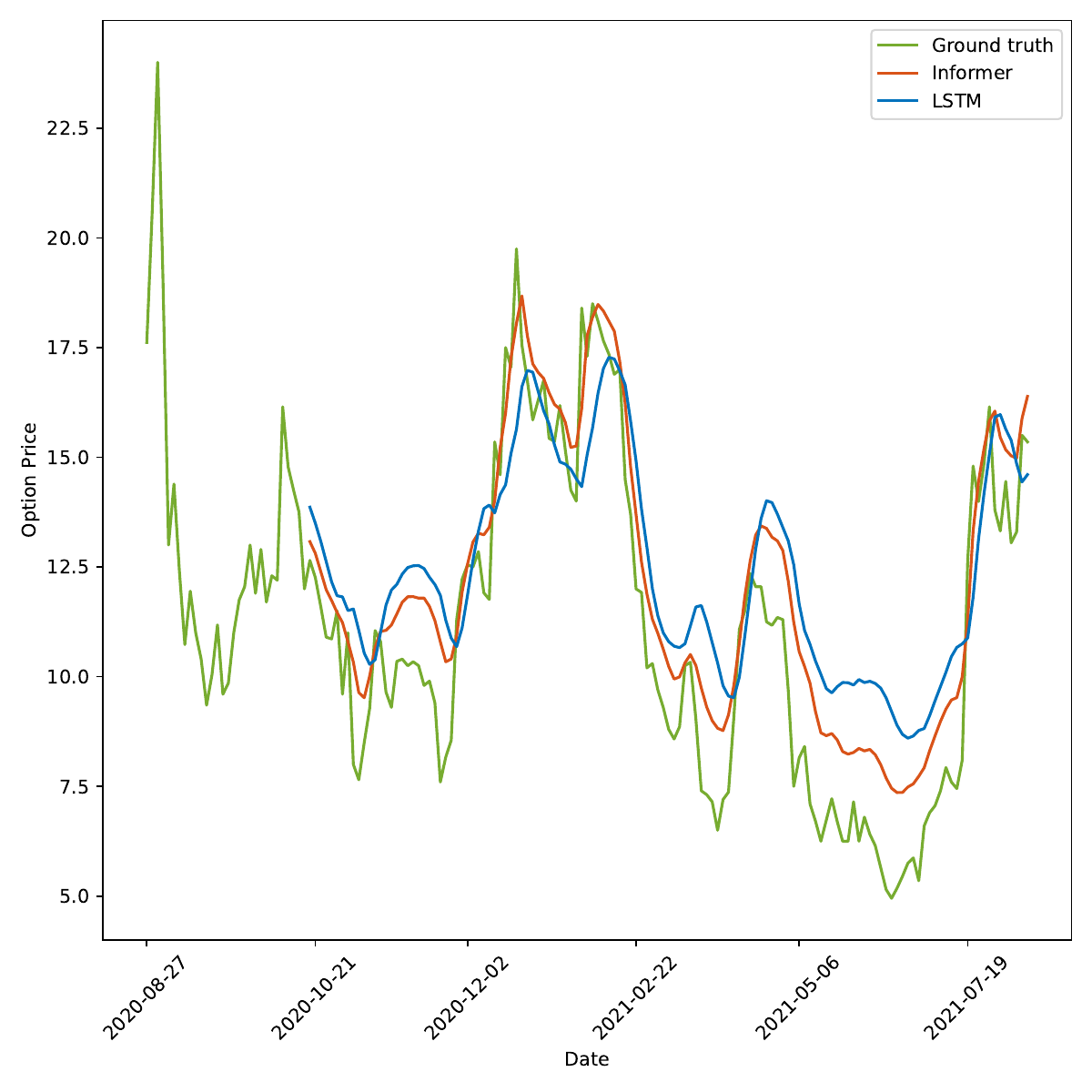}
\caption{Comparison of Informer and LSTM predictions on a longer period for an option contract.}
\label{fig:comparison_longer_period}
\end{figure}

Table~\ref{tab:overall_metrics} highlights the overall prediction metrics, Mean Absolute Error (MAE) and Root Mean Squared Error (RMSE). The Informer model achieves the lowest MAE (\(2.7145\)) and RMSE (\(3.6766\)) among all models, demonstrating its superior ability to predict option prices with high accuracy. The LSTM model, while a competitive machine-learning approach, exhibits a significantly higher MAE (\(3.9343\)) and RMSE (\(5.0373\)). Traditional models like Black-Scholes and Heston, despite their widespread use in finance, perform worse than the machine-learning-based methods. The Black-Scholes model has a slightly lower MAE (\(4.1765\)) compared to the Heston model (\(4.1282\)), but both models fail to capture complex market dynamics as effectively as the Informer. Figure~\ref{fig:comparison_longer_period} further illustrates the comparative performance of the Informer and LSTM models on a longer prediction period, highlighting the Informer's ability to track trends more closely.

\begin{table}[h!]
\centering
\caption{Final-day evaluation metrics for all models.}
\begin{tabular}{|c|c|c|}
\hline
\textbf{Model} & \textbf{DA (\%)} & \textbf{Final-Day MAE} \\
\hline
Informer & 54.43 & 2.9709 \\
LSTM & 52.19 & 4.0900 \\
Black-Scholes & 52.53 & 4.6880 \\
Heston & 51.74 & 4.6861 \\
\hline
\end{tabular}
\label{tab:final_day_metrics}
\end{table}

Table~\ref{tab:final_day_metrics} presents the final-day evaluation metrics, including Direction Accuracy (DA) and Final-Day MAE. The Informer achieves the highest DA (\(54.43\%\)) and the lowest Final-Day MAE (\(2.9709\)), showcasing its ability to predict both the direction and final value of option prices with superior precision. The LSTM model, while demonstrating a reasonable DA (\(52.19\%\)), exhibits a higher Final-Day MAE (\(4.0900\)), indicating less reliability in final price predictions. Among the traditional models, Black-Scholes performs slightly better than Heston, achieving a DA of \(52.53\%\) compared to \(51.74\%\), but both models have significantly higher Final-Day MAE values, exceeding \(4.68\).

\begin{table}[h!]
\centering
\caption{Performance of the trading strategy for Apple options based on final cumulative net value.}
\begin{tabular}{|c|c|}
\hline
\textbf{Model} & \textbf{Net Value} \\
\hline
Informer & 1.30 \\
LSTM & 1.21 \\
Heston & 1.15 \\
Black-Scholes & 1.14 \\
\hline
\end{tabular}
\label{tab:trading_performance}
\end{table}

\begin{figure}[h!]
\centering
\includegraphics[width=0.48\textwidth]{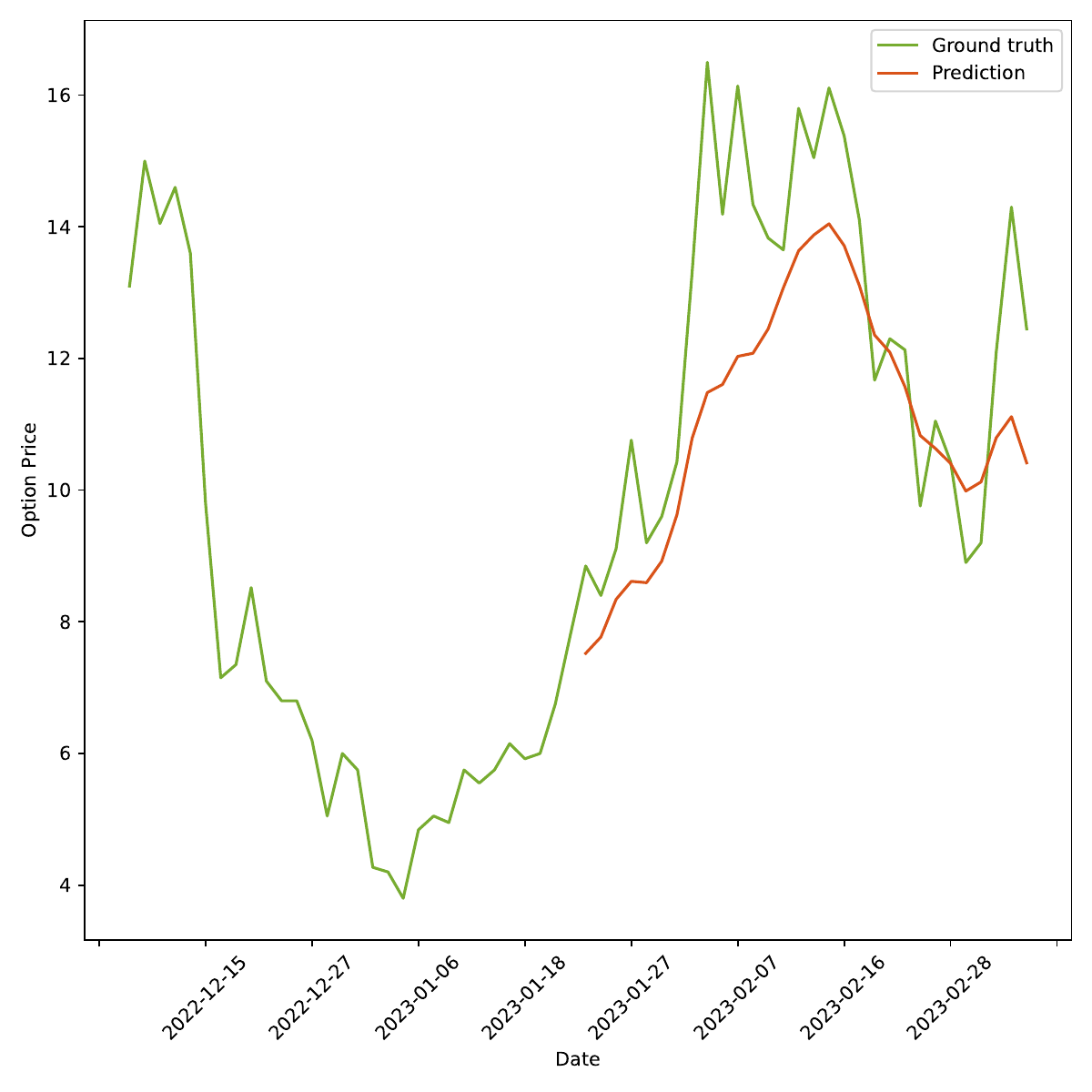}
\caption{Example of upward trend Informer prediction for one option contract.}
\label{fig:upward_trend}
\end{figure}

In trading performance, the Informer achieved the highest cumulative net value (NV), outperforming all models, as shown in Table \ref{tab:trading_performance}. With a final NV of 1.30, the Informer model demonstrates its superior ability to generate profitable trading strategies by accurately predicting directional movements over a 30-day horizon. The LSTM model follows with an NV of 1.21, while the traditional models, Heston and Black-Scholes, lag slightly behind with NVs of 1.15 and 1.14, respectively.

\begin{figure}[h!]
\centering
\includegraphics[width=0.48\textwidth]{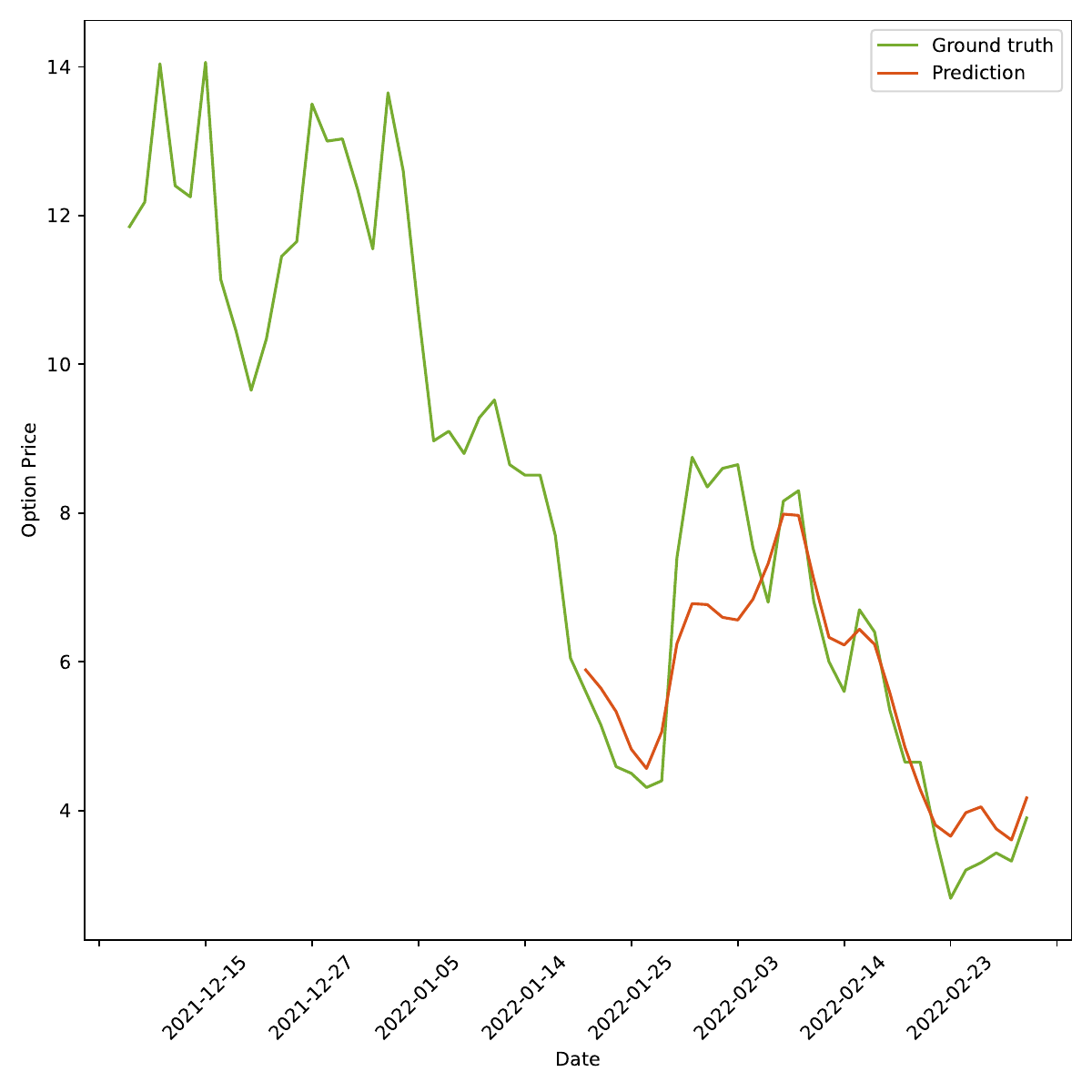}
\caption{Example of downward trend Informer prediction for one option contract.}
\label{fig:downward_trend}
\end{figure}

\begin{figure}[h!]
\centering
\includegraphics[width=0.48\textwidth]{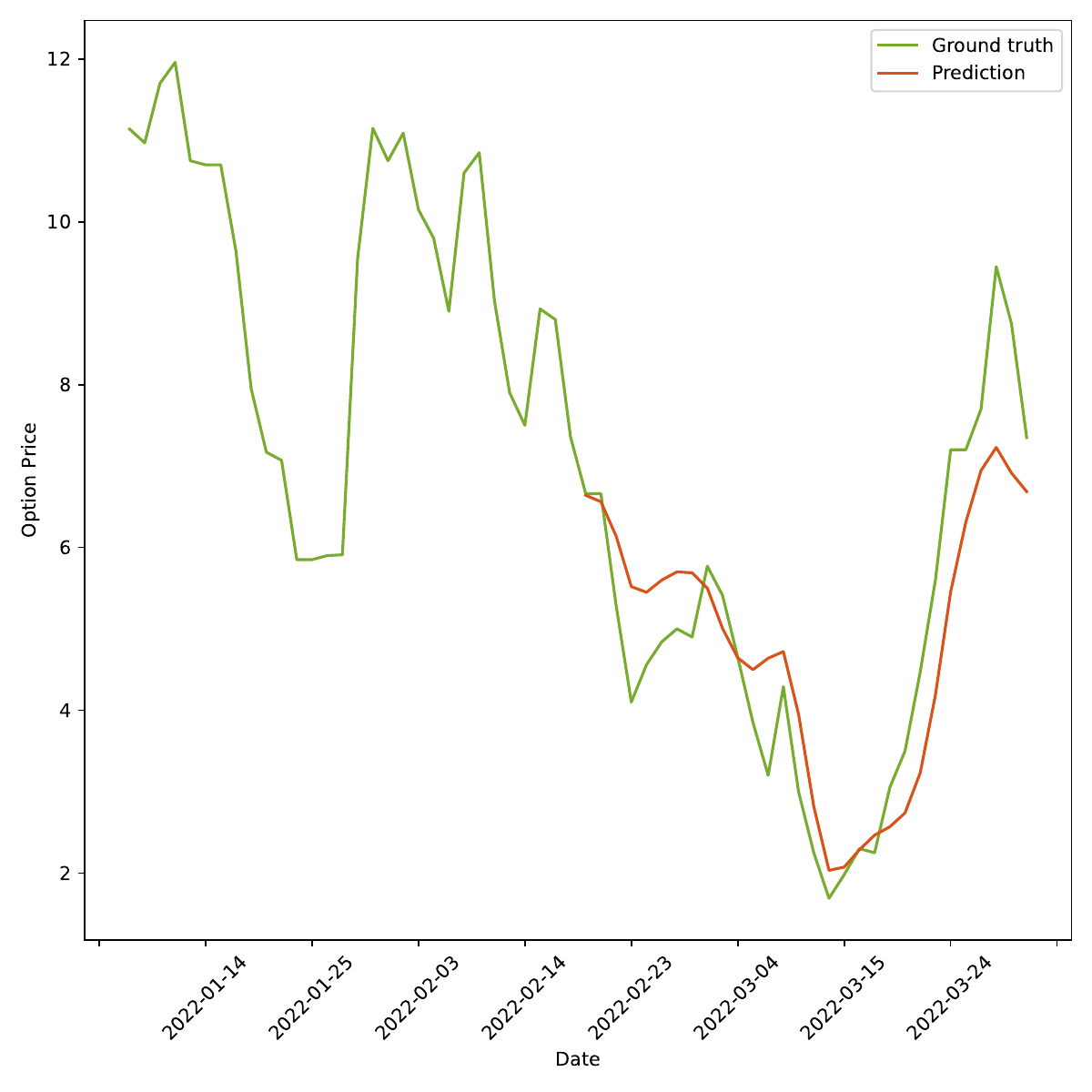}
\caption{Example of mixed trend Informer prediction for one option contract.}
\label{fig:mixed_trend}
\end{figure}

As we can see on Figures \ref{fig:upward_trend} to \ref{fig:mixed_trend} the Informer's predictions remain stable across different trend types—upward, downward, and mixed. This stability highlights the potential of the Informer model as a valuable tool for investors, providing reliable insights to navigate diverse market conditions effectively.

%% file: Conclusion.tex
\section{Conclusion and Future Work}
Our research demonstrates that the Informer model, with its specialized attention mechanisms and generative-style decoder, outperforms traditional models like Black-Scholes and Heston, as well as recurrent neural networks such as LSTM, in predicting option prices and capturing long-term dependencies in financial data. The Informer not only achieved the lowest MAE and RMSE across all tested models but also generated the highest cumulative net value in trading evaluations, outperforming all other models and demonstrating its practical value in optimizing trading strategies.

This paper contributes to the field of option pricing by implementing the Informer model for option trading and evaluating its performance against other established models.

This study demonstrates the potential of the Informer model in enhancing option pricing predictions, yet there are several avenues for further exploration. Future work could involve incorporating reinforcement learning (RL) strategies to dynamically adjust trading decisions based on model predictions \cite{Szydlowski2024}, improving adaptability in real-time trading environments. Additionally, applying the Informer architecture within a broader portfolio management framework could reveal insights into its effectiveness in balancing risk and return across diverse financial instruments. Another promising direction would be to test and refine trading strategies based on model outputs, such as mean-reversion or momentum-based approaches, to assess the practical profitability and robustness of Informer in real-world trading applications \cite{paclic_elliottagents}.